\documentclass[useAMS,usenatbib,a4paper]{mn2e}
\usepackage{graphicx}
\usepackage{amsmath}
\usepackage{amssymb}

\def\e{{\rm e}}

\newcommand{\be}{\begin{equation}}
\newcommand{\ee}{\end{equation}}
\newcommand{\bea}{\begin{eqnarray}}
\newcommand{\eea}{\end{eqnarray}}
\newcommand{\bg}{\begin{gather}}
\newcommand{\eg}{\end{gather}}
\newcommand{\bseq}{\begin{subequations}}
\newcommand{\eseq}{\end{subequations}}

\renewcommand{\ln}{\mathop{\rm ln}\nolimits}

\def\gr{$\gamma$-ray} 
\def\pks{PKS 2155-304}
\def\e{{\rm e}}

\begin{document}

\title{Measuring parameters of AGN central engines with 
very high energy \gr\ flares}

\author[Neronov et al.]{A.~Neronov$^{1,2}$, D.~Semikoz$^{3,4}$,
  S.~Sibiryakov$^{5,4}$\\ 
$^{1}$INTEGRAL Science Data Center, Chemin d'\'Ecogia 16, 1290
Versoix, Switzerland\\ 
$^{2}$Geneva Observatory, 51 ch. des Maillettes, CH-1290 Sauverny, Switzerland\\
$^3$APC, 10, rue Alice Domon et Leonie Duquet, 75205 Paris, France\\
$^4$Institute for Nuclear Research of the Russian Academy of
  Sciences,  60th October Anniversary prospect 7a, Moscow 117312, Russia\\
$^5$Theory Group, Physics Department, CERN, CH-1211 Geneva 23,
  Switzerland}
\date{Received $<$date$>$  ; in original form  $<$date$>$ }
\pagerange{\pageref{firstpage}--\pageref{lastpage}} 
\maketitle
\label{firstpage}

\begin{abstract}
  We discuss a "compact source" model of very high energy (VHE) 
emission from blazars in
  which the variability time is determined by the blazar central
  engine. In this model electron or proton acceleration close to the
  supermassive black hole is followed by the development of
  electromagnetic cascade in a radiatively inefficient accretion flow.
  Assuming such a model for the TeV blazar \pks, we show that the
  variability properties of the TeV \gr\ signal observed during a
  bright flare from this source, such as the minimal variability time
  scale and the recurrence period of the sub-flares, constrain the
  mass and the angular momentum of the supermassive black hole.
\end{abstract}


\section{Introduction}

Recent observation of fast variability of TeV \gr\ emission from
several TeV blazars \citep{aharonian07,albert07} challenges the
conventional model in which the TeV \gr s are supposed to be produced
at large distances from the blazar central engine, the supermassive
black hole. Within this conventional model the \gr\ emitting blobs
are assumed to travel to the parsec-scale distances along the AGN jet
before emitting in the TeV energy band; it is believed that 
otherwise the \gr\
emission would be strongly absorbed in the accretion flow (see
e.g. \citet{1996MNRAS.280...67G}). Since the radiative cooling time
of the TeV emitting electrons is typically shorter than the time of
propagation from the central engine to the TeV emission region, it is
usually assumed that the TeV emitting particles are produced via shock
acceleration locally in the emission region, rather than in the AGN
central engine.

At the same time, the observed short variability time scales $t_{\rm
  var}\sim$ a few minutes
indicate that the TeV emission comes from very compact regions
having the size in the comoving frame
$\Delta x'\lesssim \delta(1+z)^{-1}ct_{\rm var}$, where
$\delta$ is the bulk Doppler factor and $z$ is the source redshift.
This implies that in the static frame the longitudinal size of the 
TeV
emitting region is 
\begin{equation}
\label{rlab}
\Delta x=\frac{\Delta x'}{\Gamma}\lesssim (1+z)^{-1}ct_{\rm var}\simeq 
6\times 10^{12}(1+z)^{-1}\left[\frac{t_{\rm
      var}}{200\mbox{ s}}\right]\mbox{cm}\;. 
\end{equation}
where we assume that the bulk Lorentz factor $\Gamma\sim \delta$. This is comparable to the minimal possible scale set up by
the gravitational radius of the central supermassive black hole
\be
\label{Rg}
R_g=GM_{\rm BH}/c^2\simeq 1.5\times
10^{12}\left[M_{\rm BH}/10^{7}M_\odot\right]\mbox{ cm}\;. 
\ee
Even if one assumes that the TeV emitting plasma blobs are produced
close to the black hole, which would explain their initially small
size, it is not clear how the blobs propagating downstream the
relativistic jet can retain this size up to large distances,  unless they
have unreasonably large bulk Lorentz factors. 
This
problem has recently led to a suggestion \citep{begelman07} that the
TeV flares may be not triggered by the black hole but rather are
results of enhanced emission intrinsic to the jet.

However even in that case, to explain the
observed rapid variability, the TeV \gr\ emitting blobs have to
travel with quite large bulk Lorentz factors, 
$\Gamma\gg 1$. Such 
Lorentz factors are in contradiction with
the radio 
observations of the motion of hot spots
in the parsec-scale  jets.  Moderate apparent speeds of the blobs of the parsec-scale jets, revealed by radio observations, combined with an estimate of the number of parent objects of TeV blazars, would give  much smaller values of the bulk Lorentz factors, $\Gamma\sim 1$ \citep{henry06}. For example, to explain
the fast variability of the July 2006 TeV flare of  
\pks\ \citep{aharonian07} the bulk Lorentz factor required by the
mechanism of TeV emission in the parsec-scale jet
should be $\Gamma>50$~\citep{aharonian07,begelman07}.  At the same time, 
 the direct observations  of the apparent velocity 
of hot spots in \pks\ 
jet at the projected distance $(1\div 2)$ parsecs give the value 
$v_{\rm app}=(0.9\pm 0.3)c$  \citep{Piner:2008ju}. 
Assuming that the viewing angle is not too small, 
$\theta\gtrsim 1^\circ$,
this yields
$\Gamma\lesssim 10$ at the distance of a few tens of parsecs from the
central engine.

Both the problems of the fast variability and of the small observed
Lorentz factors at parsec distances could be naturally resolved if the
site of the VHE \gr\ production is located closer to the AGN central
engine, at significantly sub-parsec distances. If the VHE emitting
region is moving relativistically toward the observer with a bulk
Lorenz factor $\Gamma$, the variability time scale limits the distance $R$
of the \gr\ production site from the central engine (see e.g. \citet{celotti}),
\begin{equation}
R\sim \Delta x\,\Gamma^2\le 1.5\times 10^{16}(1+z)^{-1}\left[\frac{t_{\rm
      var}}{200\mbox{ s}}\right]\left[\frac{\Gamma}{50}\right]^2\mbox{ cm.}
\end{equation}
An immediate difficulty is, however, that at such distances the
accretion flow onto the black hole can be opaque to the \gr s
(see e.g. \citet{blandford95}).

The problem of opacity of the compact source does not arise in the
case of low-luminosity AGNs that accrete at significantly
sub-Eddington rates \citep{celotti}. In these sources the accretion
flow is described within the framework of the radiatively inefficient
accretion flow (RIAF) models
\citep{rees82,narayan94,narayan95,narayan02} in which most of the
gravitational energy extracted from the accreted matter is converted
into internal energy, rather than into radiation. The
possibility of escape of the VHE \gr s from the vicinity of the AGN
central engine is best illustrated by the nearby low-luminosity radio
galaxy M87, which was recently found to be a source of the variable TeV
\gr\ emission \citep{aharonian03,aharonian06,albert08}, most probably coming
from a compact source \citep{neronov07,aharonian08a}.

In the compact source model the VHE \gr\ emission is triggered by 
high-energy particles
accelerated close to the black hole via one of the possible
mechanisms (see
e.g. \citet{lovelace76,Lesch1992,kardashev95,Bednarek:1998jq,neronov02,
neronov02a,neronov04,
Rieger:2007tt,neronov08}). In this case the spectral and timing
characteristics of the VHE emission are directly linked to the physics
of the processes taking place close to the supermassive black hole,
which naturally explains the variability of the signal on the shortest
possible time scale.

Within the AGN unification scheme, the TeV blazars (high-energy peaked
BL Lacs) are assumed to be the beamed versions of the low-luminosity
radio galaxies similar to M87
\citep{browne83,giroletti04,giroletti06}.  Since the only difference
between the TeV blazars and the low-luminosity radio galaxies is their
orientation with respect to the line of sight, the compact source
model can be applicable also in the case of TeV blazars.

In what follows we adopt this point of view and develop a compact
source model of high-energy activity of TeV blazars. We demonstrate
that within such compact source model the characteristics of the
fast-variable VHE emission can be used to constrain the parameters of
the AGN central engine, in particular, the density and luminosity of
the accretion flow, the black hole mass and spin. We illustrate this
possibility on the particular example of the bright \pks\ 
flare detected by the HESS telescope in July 2006 \citep{aharonian07}.
This flare consists of a number of well-pronounced sub-flares which
exhibit quasi-periodic recurrence.
We show that the rise time and the recurrence period of the sub-flares
can be directly related to the light-crossing time and to the period
of rotation over the last stable orbit around a $M_{\rm BH}\sim
10^7M_\odot$ black hole\footnote{This mass estimate is different from
  the value $\sim 10^9 M_\odot$ quoted by \citet{aharonian07}. 
We will comment on
this discrepancy in Sec.~\ref{sec:3.3}.}.
 
The paper is organized as follows.  In Sec.~\ref{IR} we discuss the
qualitative features of the model, including particle acceleration and
propagation through the RIAF environment. The possibility of a new
interpretation of the observational data within such a model is
demonstrated in Sec.~\ref{TIMING} where we find the constraints
on the black hole mass and angular momentum imposed by the timing
analysis of the bright TeV flare of \pks.
We summarize our results in Sec.~\ref{CONCL}.

\section{Compact source model of the TeV blazars}
\label{IR}

In the compact source model the VHE \gr\ emission region is assumed
to be situated in the vicinity of the AGN central engine, rather than at
parsec-scale distances. This fact implies two main differences of this
model from a generic "relativistically moving blob" model of VHE
emission from blazars. First, the high-energy particles responsible
for the VHE emission can be injected into the \gr\ emission region by
the AGN central engine, rather than only by a process (shock
acceleration) intrinsic to the blob. Second, the characteristics of
the \gr\ emission are determined not only by the intrinsic properties
of the blob, but also by the effects of propagation of the high-energy
particles through the matter and radiation environment created by the
accretion flow. However, as we discuss below, the compact source model
to some extent includes the blob model: the
electromagnetic cascade, which develops as a result of the 
propagation of
high-energy particles through the accretion flow environment, leads to
the formation of a relativistically moving blob of secondary cascade
particles. In the following sub-sections we consider general features
of particle acceleration and propagation in the compact source model
of TeV blazars.

\subsection{Particle acceleration and "direct" \gr\ emission from the
  acceleration region}

A number of mechanisms of particle acceleration in the vicinity of the
central black hole have been proposed in the literature (see
e.g. \citet{lovelace76,Lesch1992,kardashev95,
Bednarek:1998jq,neronov02,neronov02a,neronov04,Rieger:2007tt,neronov08}). In
this sub-section we summarize some common features of the acceleration
models which are related to the fact that the acceleration proceeds in a
compact region of the size comparable to the black hole horizon and
are independent of the details of the acceleration
mechanism.

A conventional dimensional estimate of maximal possible energies of
particles of charge $e$ accelerated in a region of the size $R\sim
R_g$ with the magnetic field $B$ is 
\be
E_{\rm max}= \kappa eBR_g\simeq
10^{19}\kappa \left[\frac{B}{10^4\, \mbox{G}}\right] \left[\frac{M_{\rm
    BH}}{10^{7}M_\odot}\right]\mbox{ eV}\;,
\ee 
where $\kappa\le 1$ is the
efficiency of a particular acceleration mechanism. This maximal energy
is, however, not always achieved because of the strong energy losses
experienced by the accelerated particles. The "minimal" energy loss
channel is the loss on the curvature radiation
\citep{levinson00,neronov05,neronov08}, which limits the particle
energies to
\be
 \frac{E_{\rm cur}}{mc^2}\le
\left[\frac{3 R_g^2 \kappa B}{2e}\right]^{1/4}
\!\!\simeq
3\times 10^9\kappa^{1/4}
\left[\frac{M_{\rm BH}}{10^{7}M_\odot}\right]^{1/2}
\left[\frac{B}{10^4\mbox{ G}}\right]^{1/4} 
\label{curv_cut}
\ee
where $m$ is the particle mass. Here we normalize the magnetic field to
the value $10^4\mbox{ G}$ typical for the central engine of an AGN
with a $10^7 M_\odot$ black hole. 
Additional energy losses, caused by
the interactions of the accelerated particles with the matter and
radiation backgrounds produced by the accretion flow (see
Sec. \ref{riaf} below), lead to further reduction of the maximal
attainable energy. If the magnetic field in the acceleration region is
not ordered, the energies of the accelerated particles are reduced
because of the strong synchrotron energy loss.

Radiative energy losses which accompany particle acceleration
(curvature, synchrotron, inverse Compton) result in the \gr\ emission
directly from the acceleration region. If the resulting \gr s are not
completely absorbed during their propagation through the photon
background created by the accretion flow, this "direct" \gr\ emission
can, in principle, provide an observable signature of the compact
source model. We will return to this point in Sec. \ref{blobs}.

Since the maximal energies of particles are determined by the balance
of the acceleration and energy loss rates, all the work done by the
electric field is dissipated via the available energy loss
channels. For each charged particle the energy loss rate is about the
acceleration rate, $dE/dt \sim e\kappa Bc$, where we assume that the
accelerating electric field strength is $\sim \kappa B$. The maximal
possible density of particles in the acceleration region, $n_q\sim
\kappa B/eR$, is determined by the condition that a charge
redistribution cannot screen the electric field. One estimates 
the total power extracted from the acceleration region by
multiplying the energy loss rate of each particle
on the maximal possible particle density and on the volume of the
acceleration region, $V\sim R_g^3$; this yields
\begin{eqnarray} 
\label{power} 
P_{\rm tot}&\simeq &n_qR_g^3(dE/dt) \\&\simeq&  
5\times 10^{42}\kappa^2\left[\frac{M_{\rm BH}}{10^7M_\odot}\right]^2\left[\frac{B}{10^4\mbox{ G}}\right]^2 
\mbox{ erg/s.}\nonumber 
\end{eqnarray} 
Comparing this estimate to the apparent luminosity of a bright TeV
flare (found under the assumption that the radiaiton is isotropic), $L_{\rm iso}\simeq 10^{46}\mathrm{erg/s}$, one concludes that
the energetics of such a flare is compatible with the compact
source model if the TeV emission is beamed into the solid angle 
$\Omega/4\pi\lesssim 10^{-3}$. The emission from the compact source is expected to be anisotropic, with the direction of the beam of high-energy particles and/or photons set up by the direction of the magnetic field. 

A fraction of the work done by the accelerating electric field is
carried away by the flux of accelerated particles (a relativistic
wind). In the case of electrons, this fraction is usually negligible.
In the case of protons or heavy nuclei the power of
the relativistic particle wind, $P_p$, can be estimated as
\begin{align}
\label{power1}
&P_p\sim 
n_qR_g^2c\cdot\min\{E_{\rm cur},E_{\rm max}\}
\\
&\simeq  \left\{
\begin{array}{ll}
\!\!10^{42}\kappa^{5/4}\left[\frac{M_{\rm
      BH}}{10^7M_\odot}\right]^{3/2}\left[\frac{B}{10^4\mbox{
      G}}\right]^{5/4}\frac{\mbox{ erg}}{\mbox{s}}\,, & E_{\rm
  cur}<E_{\rm max}\\
5\times 10^{42}\kappa^2\left[\frac{M_{\rm
      BH}}{10^7M_\odot}\right]^2\left[\frac{B}{10^4\mbox{ G}}\right]^2
\frac{\mbox{ erg}}{\mbox{s}}\,,& E_{\rm cur}>E_{\rm max}
\end{array}
\right.\nonumber
\end{align} 
This order-of-magnitude estimate does not take into account a possible
special geometry of the accelerating field. If, for instance, particles are
accelerated in parallel magnetic and electric fields in the polar
cap regions of black hole magnetospheres, the curvature radii of
particle trajectories can be somewhat larger than $R_g$ and,
respectively, the maximal attainable energies and the power of the
particle beam can be higher \citep{neronov08}.
 
The high-energy particles and radiation emitted from the acceleration
region are injected into the accretion flow. The form of the
\gr\ signal produced by the interaction of the accelerated
particles with the radiation coming from the accretion flow depends
sensitively on the geometry of the central engine. The precise
determination of this signal should involve detailed numerical
modeling which is beyond the scope of the present paper. Below we
limit ourselves to order-of-magnitude estimates which 
highlight the qualitative features of the problem.

\subsection{Radiatively inefficient accretion flow}
\label{riaf}

In this subsection we summarize the general properties of the
radiation produced by the accretion flow. As it is mentioned in the
Introduction, the TeV blazars belong to the class of low-luminosity
radio galaxies in which the accretion is conventionally described in
the framework of RIAF models (see e.g. \citet{narayan02} and
references therein) relevant for sources in which the accretion rate
$\dot M$ is small in the Eddington units,
\begin{equation}
  \dot M\ll \dot M_{\rm Edd}\equiv\frac{L_{\rm Edd}}{0.1c^2}\simeq
  0.2\left[\frac{M_{\rm BH}}{10^7M_\odot}\right]\ M_\odot\mbox{/yr}\;, 
\end{equation}
where $\dot M_{\rm Edd}$ is the Eddington accretion rate.

The low radiative efficiency of the accretion flow is attributed to
the low matter density which leads to inefficient cooling of electrons
and ions in the accretion flow. As a result, a large fraction of the
released gravitational energy, instead of being dissipated in the form
of radiation, is stored in the internal energy of matter. This energy
either disappears under the horizon or is ejected in an outflow.

Large internal energy of matter does not allow formation of a
geometrically thin, optically thick accretion disk. Instead, the
accreting matter forms around the BH a hot optically thin gaseous torus
\citep{rees82}. In RIAF models the accretion torus typically consists
of two-temperature plasma with both electrons and ions being mildly
relativistic. The radiation of the plasma is entirely produced by
electrons via three main mechanisms: synchrotron
radiation, comptonization of the latter, and bremsstrahlung.

The synchrotron radiation contributes to the far infrared part of the
spectrum. The characteristic synchrotron emission energy is
\begin{equation}
\epsilon_{\rm synch}=\frac{\hbar e B}{m_ec}\langle\gamma_e^2\rangle
\simeq
5\times 10^{-3}\left[\frac{B}{10^{4}\mbox{ G}}\right]
\left[\frac{T_e}{1\mbox{ MeV}}\right]^2\mbox{eV}\;,
\end{equation}
where $B$ is the magnetic field in the inner part of the accretion
torus, $m_e$ is the electron mass, $<\gamma_e^2>$ is the mean-square
$\gamma$-factor of electrons, $T_e$ is the electron
temperature, and we have used that for mildly relativistic electrons
\be
\label{gammae}
\langle\gamma_e^2\rangle\sim 10 \left(\frac{T_e}{m_e c^2}\right)^2\;.
\ee 
One of the parameters of the RIAF models is the ratio $\beta$ of
the gas pressure $p_{\rm gas}=n m_pv_{\rm th}^2/3$ to the magnetic
pressure $p_{\rm magn}=B^2/4\pi$. Here $n$ is the gas density, $m_p$
is the proton mass and $v_{\rm th}$ is the thermal velocity of protons
which can be estimated as the Keplerian velocity at the corresponding
distance, $v_{\rm th}(R)\approx (R_g/R)^{1/2}$. With the use of parameter $\beta$, the magnetic field is estimated as
\begin{eqnarray}
\label{eq:B}
B(R)&=&\left[\frac{4\pi}{3}\beta^{-1}n(R)m_pv_{\rm th}^2(R)\right]^{1/2}\\
&\simeq&
10^4\beta^{-1/2}\left[\frac{R_g}{R}\right]^{1/2}\left[\frac{n(R)}{10^{10}\mbox{
      cm}^{-3}}\right]^{1/2}\mbox{G}\;. 
\nonumber
\end{eqnarray}
Note that we normalize the matter density in the black hole vicinity
to the value $10^{10}{\rm cm}^{-3}$, typical for the case of a
$10^7M_\odot$ black hole in the RIAF models, cf.~\citet{narayan95}.

If the matter density changes with the distance as
$n(R)\sim R^{-\gamma}$ (in the RIAF models $1/2\le \gamma\le 3/2$
\citep{lu04})
the magnetic field decreases as $B\sim R^{-(\gamma+1)/2}$, so that the
synchrotron cooling time
\begin{equation}
t_{\rm synch}=\frac{6\pi m_ec^2}{\sigma_TB^2\gamma_e}
\simeq 1.3\left[\frac{B}{10^4\mbox{ G}}\right]^{-2}
\left[\frac{T_e}{1\mbox{ MeV}}\right]^{-1}\mbox{ s}
\end{equation}
increases with the distance as $t_{\rm synch}\sim R^{(\gamma+1)}$. 
Here $\sigma_T=0.665\times 10^{-24}$~cm$^2$ is the Thomson cross-section. The
synchrotron cooling is efficient only close to the BH horizon. Indeed,
the synchrotron cooling time grows with $R$ faster than the dynamical
time scale given by the free-fall time, 
\be
\label{ff}
t_{\rm ff}\simeq \sqrt{R^3/GM_{\rm BH}}
\ee 
Thus, the bulk of the
synchrotron emission is produced in the vicinity of the
horizon. Numerical simulations \citep{narayan02} confirm this
qualitative result. 

The synchrotron emission in the RIAF models is damped due to 
self-absorption. The self-absorption coefficient has the form
\citep{Pacholczyk}
\be
\label{alphaSA}
\alpha_{\rm SA}(\epsilon)=
\frac{\pi\hbar e^2m^2c^5n}{2\sqrt{3}T_e^3\epsilon}
I(x)\;,
\ee 
where 
\be
\label{x}
x=\frac{2m^3 c^5\epsilon}{3\hbar eBT_e^2}
\ee
and the function $I(x)$ is given explicitly in the 
appendix~\ref{App}. Equating $\alpha_{\rm SA}^{-1}$ to the size of the
synchrotron emission region, which we take to be of order of the gravitational radius,
one finds the photon energy at which the synchrotron emission becomes
optically thin
\be
\label{SA}
\epsilon_{\rm SA}\simeq 37\,\epsilon_{\rm synch}\;. 
\ee
As discussed in the appendix \ref{App}, the numerical coefficient in
this 
formula
mildly depends on the parameters of the accretion flow. We omit this
dependence in what follows. 

Since $\epsilon_{\rm SA}>\epsilon_{\rm synch}$, the maximum of the
synchrotron power is emitted at the energy $\epsilon_{\rm SA}$. To
estimate the synchrotron luminosity we approximate the spectrum by 
thermal radiation up to $\epsilon_{\rm SA}$. This yields
\be
\label{Lsynch}
L_{\rm synch}=4\pi R^2_{\rm synch}\frac{2 T_e}{c^2}
\int_0^{\epsilon_{\rm SA}}\frac{\epsilon^2 d\epsilon}{(2\pi\hbar)^3}\;,
\ee
where $R_{synch}$ is the size of the synchrotron emission
region. Substituting
$R_{\rm synch}\approx R_g$ we obtain
\be
\label{Lsynch1}
L_{\rm synch}\simeq 3.2\times 10^{39}
\!\left[\frac{M_{\rm BH}}{10^7 M_\odot}\right]^2
\!\left[\frac{B}{10^4\mbox{ G}}\right]^3
\!\left[\frac{T_e}{1\mbox{ MeV}}\right]^7
\!\mbox{erg/s}\;.
\ee 
The part of the synchrotron spectrum above $\epsilon_{SA}$ gives small
contribution to the total synchrotron power because of the exponential
cutoff in the synchrotron emission function. An immediate consequence
of this cutoff is that the bulk of the synchrotron photons have
energies $\epsilon\lesssim \epsilon_{\rm SA}\simeq 0.2\mbox{ eV}$.

The synchrotron radiation is upscattered by inverse Compton (IC) 
process
into the optical band,
\begin{equation}
\epsilon_{IC}\sim\epsilon_{\rm SA}\langle\gamma_e^2\rangle
\simeq 4\,\left[\frac{B}{10^{4}\mbox{ G}}\right]
\left[\frac{T_e}{1\mbox{ MeV}}\right]^4\mbox{eV}\;.
\end{equation}
 To estimate the luminosity of the accretion flow
in this range we note that the IC volume emissivity 
at
distance $R$ from the black hole is 
\citep{lightman}
\be
\label{dLIC}
\frac{dL_{\rm IC}}{dV}=\frac{4}{3}\sigma_Tcn(R)
\langle\gamma_e^2(R)\rangle
U_{\rm synch}(R)\;,
\ee
where $\langle\gamma_e^2(R)\rangle$ is  the
mean-square gamma-factor of electrons at this distance, and 
$U_{\rm synch}(R)\simeq L_{\rm synch}/(4\pi R^2c)$ is the energy density of the
synchrotron radiation. Integrating over the volume we obtain
\be
\label{LIC}
L_{\rm IC}\simeq \frac{4}{3} L_{\rm synch}\,
\sigma_T\int n_e(R)\langle\gamma_e^2(R)\rangle dR\;.
\ee
Depending on the radial profiles of the electron density and
temperature, the integral in (\ref{LIC}) is saturated either close to
the black hole horizon (for the radial profiles
steeper than $n(R)\langle\gamma_e^2(R)\rangle\sim R^{-1}$) 
or in a region which is much larger
than the size of the black hole. In the numerical simulations of RIAF
models one finds the typical size $R_{\rm IC}$ 
of the IC
emission region to be of order $100\,R_g$ \citep{narayan02}.  
As a crude estimate let us
consider the case when the integral in (\ref{LIC}) is saturated at the
upper limit. Then,
\be
\label{LIC1}
\begin{split}
L_{\rm IC}\simeq 0.4\times 10^{39}
&\left[\frac{L_{\rm synch}}{10^{39}\mbox{ erg/s}}\right]
\left[\frac{R_{\rm IC}}{10^{14}\mbox{ cm}}\right]
\\
&\times\bigg[\frac{n(R_{\rm IC})}{10^8\mbox{ cm}^{-3}}\bigg]
\bigg[\frac{T_e(R_{\rm IC})}{1\mbox{ MeV}}\bigg]^2
\mbox{ erg/s}\;.
\end{split}
\ee
Note that in the above estimate we took into account that 
the electron density and temperature 
at the distance $R_{\rm IC}$ are lower than near
the black hole.

Finally, we consider the bremsstrahlung radiation. 
It contributes primarily into the
 X-ray/soft \gr\  band with the total power typically 
comparable or lower than that of the synchrotron radiation. 
The bremsstrahlung emission power is proportional to $n^2$, so that
the bremsstrahlung luminosity from a region of the size $R$ is $P_{\rm
  brems}\sim n(R)^2R^3$. If the radial density falls down slower
than $n(R)\sim R^{-3/2}$ (as it is the case in typical RIAF models),
the bulk of the bremsstrahlung flux is produced at large
distances. Numerical modeling of the radiatively inefficient
accretion shows that the bremsstrahlung is produced mainly at large
distances from the BH, $R_{\rm brems}\sim 10^5\times R_g$
\citep{narayan02}.

To summarize, the properties of RIAF important for the high-energy
particle propagation are as follows. Matter distribution in a typical
RIAF is characterized by a rather shallow radial density profile
$n(R)\sim R^{-1/2}\div R^{-3/2}$, so that the central matter density
is rather low. At the same time, due to the large non-dissipated
mechanical energy of the accreting matter the magnetic field produced
by the  RIAF can be strong close to the central black hole
(\ref{eq:B}). The radiation environment created by a RIAF has an
"onion-like" structure with the infrared synchrotron emission
produced close to the black hole, infrared/optical IC
emission produced in a larger region of the size $\sim 100R_g$ and
X-ray/soft \gr\ bremsstrahlung emission produced in a region of the
size $\sim 10^5R_g$.

To conclude this section let us make the following comment. The
estimates presented above strongly depend on the values of the
parameters of the accretion flow such as electron temperature, density
and magnetic field. These characteristics vary significantly in
different RIAF models. The estimates presented in this section should
be considered as indicative. A more detailed analysis of the the
radiative background can be done using numerical simulations in each
particular RIAF model.

\subsection{Propagation of the high-energy particles through the accretion flow background}
\label{sec:absorb}

\begin{figure}
\includegraphics[width=\linewidth]{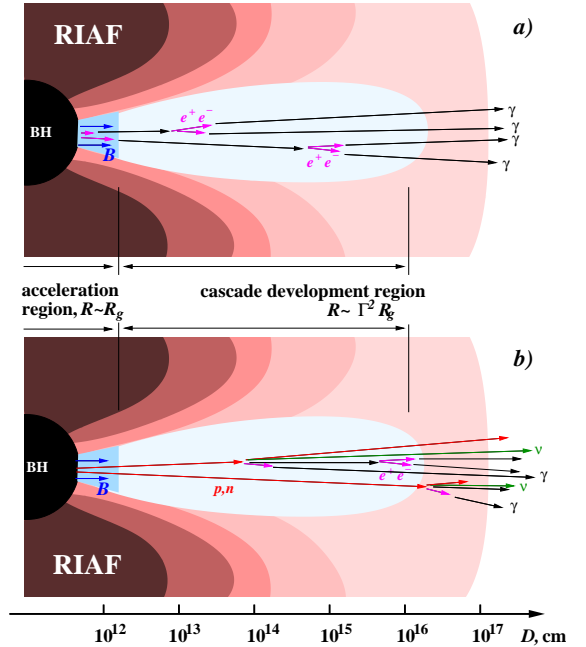}
\caption{Schematic representation of the two possible scenarios for the \gr\ emission from the compact source. Particles (electrons in the upper panel, protons in the lower panel) are initially accelerated in a compact region of the size of about $R_g$, shown as a blue-shaded region.  Particles which escape from the acceleration region initiate a cascade in the radiatively-inefficient accretion flow, at the distances $R\gg R_g$, shown as a light-blue shaded region.  }
\label{fig:cartoon}
\end{figure}

The compact particle accelerator close to the black hole produces 
emission with the total power  $P_{\rm tot}$,
Eq.~(\ref{power}). As discussed above, the partition of this
energy between the \gr\ component and the accelerated matter depends on
the type of accelerated particles, see Fig. \ref{fig:cartoon}. If the accelerated particles are
electrons the power goes completely into \gr s, while in the case of
the proton acceleration a significant part of $P_{\rm tot}$ can remain in
the high-energy particle beam, see Eq.~(\ref{power1}).  
The primary \gr s and protons propagate through and interact
with the matter and radiation background created by
RIAF. Interactions of high-energy particles give rise to
electromagnetic cascades that redistribute the power initially
contained in the highest energy particles to lower energy bands.  
Since
the rates of interactions of electrons and \gr s are significantly
different from those of protons, we consider these two cases
separately.

\subsubsection{Electrons and \gr s}

The highest energy \gr s  produce pairs in interactions with the soft
photon background inside the compact source. The cross-section of
photon--photon pair-production depends on the center-of-mass energy of
colliding photons, 
\begin{equation}
\label{s}
s=E\epsilon(1-\cos\theta)/2m_e^2\,,
\end{equation}
where $E$ and $\epsilon$ are the energies of the photons, and $\theta$ is
the collision angle. Starting from the threshold at
$s=1$ the cross-section rapidly increases achieving the maximum
$\sigma_{\gamma\gamma}\approx 1.3\times 10^{-25}\,{\rm cm}^2$ at
$s\approx 4$, and then decreases as $s^{-1}\ln s$. Because of
relatively narrow distribution of $\sigma_{\gamma\gamma}(s)$, \gr s
interact most efficiently with the optical/infrared background photons
of energy $\epsilon\simeq 1(E_\gamma/1\mbox{ TeV})^{-1}$~eV.
This interaction leads to absorption of the \gr s. In order for the
VHE emission to escape from the vicinity of the black hole the optical
depth of this process should be less than one. 
Let us consider the contributions of the various backgrounds produced
by RIAF to the optical depth, case by case.

We start with the synchrotron background. This background is produced
in the vicinity of the black hole, $R_{\rm synch}\approx R_g$. 
Consequently, its interaction with the VHE emission strongly
depends on the geometry of the inner part of the accretion disc
and of the acceleration region. The situation is still more
complicated because of the exponential dependence of the density of
the synchrotron
background photons on energy \citep{Pacholczyk},
\begin{equation}
\label{neps}
n_{\rm synch}(\epsilon)= n_{\rm synch}(\epsilon_{\rm SA})
I(x)/I(x_{\rm SA})\;,
~~~~\epsilon>\epsilon_{\rm SA}\;,
\end{equation}
where the function 
$I(x)$ is given in Eq.~(\ref{Iasymp}) and $x$ is defined
in Eq.~(\ref{x}). 
The density $n_{\rm synch}(\epsilon_{\rm SA})$ at the maximum of the
synchrotron spectrum 
can be estimated as
\begin{eqnarray}
&&n_{\rm synch}(\epsilon_{\rm SA})=\frac{L_{\rm
  synch}}{4\pi R_{\rm synch}^2\epsilon_{\rm SA} c}
\\ &&\simeq 10^{17}\!\left[
\frac{L_{\rm synch}}{10^{40}\mbox{ erg/s}}\right]
\!\left[\frac{R_{\rm synch}}{10^{12}\mbox{ cm}}\right]^{-2}
\!\left[\frac{
\epsilon_{\rm SA}}{0.2\mbox{ eV}}\right]^{-1}\!\!\mbox{cm}^{-3}\nonumber
\end{eqnarray}
Using Eq.~(\ref{neps}) one finds that because of the sharp cut-off in the synchrotron spectrum
above the energy $\epsilon_{\rm SA}\sim 0.2$~eV,
\gr s with energy $E_\gamma\sim 1$ TeV escape through the
synchrotron background, $\tau_{\gamma\gamma}(E_\gamma=1{\rm TeV})\sim 1$, if $R_{\rm
  synch}\sim 10^{12}{\rm cm}$ and $L_{\rm synch}\lesssim 10^{40}{\rm erg/s}$.
On the other hand, our crude estimates indicate that  the spectrum of the escaping \gr s 
should be sharply cut-off at the energy $E_\gamma\simeq 5\left[\epsilon_{\rm SA}/0.2\mbox{ eV}\right]^{-1}$~TeV, since  $\tau_{\gamma\gamma}$ rapidly grows to $\tau_{\gamma\gamma}\gg 1$ at this energy. One should note, however, that the details of the behaviour of the spectrum close to the cut-off strongly depend  on the details of the spatial distribution of the synchrotron emission. Since the threshold of the pair production, $s=1$ (see Eq. (\ref{s})), depends on the angle $\theta$ between the \gr\ and synchrotron photon velocities,  the anisotropy of the synchrotron emission can result in a shift of the cut-off in the \gr\ spectrum toward higher energies. 

VHE \gr s produced outside of the synchrotron emission region, or
those which escape from it, pass through the IC
background with the photon density:
\begin{eqnarray}
&&n_{\rm IC}=\frac{L_{\rm
  IC}}{4\pi R_{\rm IC}^2\epsilon c}
\\ &&\simeq 1.7\times 10^{11}\!\left[
\frac{L_{\rm IC}}{10^{39}\mbox{ erg/s}}\right]
\!\left[\frac{R_{\rm IC}}{10^{14}\mbox{ cm}}\right]^{-2}
\!\left[\frac{
\epsilon}{1\mbox{ eV}}\right]^{-1}\!\!\mbox{cm}^{-3}\nonumber
\end{eqnarray}
The optical depth of this background for the
TeV \gr s can be estimated as 
\be
\label{eq:tau_IC}
\begin{split}
&\tau_{\gamma\gamma}^{\rm IC}\left(E_\gamma\right)=
\sigma_{\gamma\gamma}n_{\rm IC}R_{\rm IC}
\\ &\simeq 2.2 \left[
\frac{L_{\rm IC}}{10^{39}\mbox{ erg/s}}\right]
\left[\frac{R_{\rm IC}}{10^{14}\mbox{ cm}}\right]^{-1}
\left[\frac{E_\gamma}{1\mbox{ TeV}}\right]
\end{split}
\ee 
Thus, the inner part of the accretion flow is transparent for the TeV \gr s if 
\begin{equation}
\label{eq:L}
L_{\rm IC}\lesssim
0.5\times 10^{39}\left[\frac{R_{\rm IC}}{10^{14}\mbox{ cm}}\right]
\left[\frac{E_\gamma}{1\mbox{ TeV}}\right]^{-1}\mbox{ erg/s}\;.
\end{equation}
Finally, the background of the bremsstrahlung radiation does not
affect propagation of the high-energy \gr s. This is due to two
reasons: the low density of the corresponding photons and their
relatively weak interaction with the VHE \gr s because of the $1/s$
suppression of the photon-photon cross-section.
 
Thus, in the case when the luminosity of the
accretion flow is as low as $L_{\rm acc}\lesssim 10^{40}{\rm erg/s}$,
the VHE \gr\ emission
can
originate directly from the vicinity of the black hole. It can be
e.g. the direct synchrotron/curvature emission which accompanies the
acceleration process. This possibility is illustrated schematically in Fig. 1 (upper panel). Otherwise, if the synchrotron luminosity of RIAF
is significantly above $10^{40}\mathrm{erg/s}$ and/or the IC
luminosity exceeds 
the limit (\ref{eq:L}), the VHE \gr\ emission must be
produced outside the corresponding radiation regions. This is possible
if the primary accelerated particles are protons.

\subsubsection{Protons}

Protons with energies above the threshold
\begin{equation}
E_{\rm th} = \frac{m_\pi(m_\pi+2 m_P)}{4\epsilon} \sim  8\times 10^{16}
\left[\frac{\epsilon}{1\mbox{ eV}}\right]^{-1}~\mbox{eV} 
\label{threshold}
\end{equation}
loose energy in the
interactions with the infrared photons of energy $\epsilon$ via
pion production. Near the threshold, the cross-section is dominated by the 
single
pion production resonance  $\sigma_{p\gamma}\sim 6 \times
10^{-28}$~cm$^2$; in this regime 
proton gives only 
$20\%$ of its energy to the pion. At higher proton energies the  
photo-pion production cross-section decreases down to 
$\sigma_{p\gamma}\sim 10^{-28}$~cm$^2$, while the
proton energy loss in every interaction increases up to 
$50\%$.
Thus in both cases protons lose most
of their energy at similar distance.

For high synchrotron background, $L_{\rm synch}\sim 10^{42}{\rm
  erg/s}$, protons with energies $E_p>10^{18}$ eV cannot
escape from the acceleration region,
\begin{equation}
\begin{split}
&\tau_{p\gamma}^{\rm synch}\left(E_p>10^{18}\mbox{eV}\right)=
\sigma_{p\gamma}n_{\rm synch}R_{\rm synch}
\\ &\sim 10^{3} \left[
\frac{L_{\rm synch}}{10^{42}\mbox{ erg/s}}\right]
\left[\frac{R_{\rm synch}}{10^{12}\mbox{ cm}}\right]^{-1}
\!\left[\frac{\epsilon}{0.2\mbox{ eV}}\right]^{-1}\!\!\;.
\end{split}
\label{tau_p_gamma}
\end{equation}
On the other hand, protons with energy $E_p < 10^{18}$eV 
interact only with exponential tail of 
the synchrotron emission and escape from this region.

During the
propagation through the inverse Compton emission region at the
distance scales $R\sim 100R_g$, protons interact with the 
inverse Compton  photon background, so that the
optical depth is
\be
\label{eq:tau_p_IC}
\begin{split}
&\tau_{p\gamma}^{IC}=
\sigma_{p\gamma}n_{\rm IC}R_{\rm IC}
\\ &\simeq 0.2 \left[
\frac{L_{\rm IC}}{10^{41}\mbox{ erg/s}}\right]
\left[\frac{R_{\rm IC}}{10^{14}\mbox{ cm}}\right]^{-1}
\!\left[\frac{\epsilon}{1\mbox{ eV}}\right]^{-1}\!\!
\end{split}
\ee 
The power of the absorbed proton flux is converted into the products
of pion decays: neutrinos, \gr s and electrons of the energies $\sim
0.1 E_p$. Obviously, the neutrinos freely escape from the production
region. 

Naively, from the above discussion of propagation of  \gr s,
one would expect that \gr s are not able to escape from such a dense
IC background. However, this is
incorrect. The point is that the energies of the \gr s produced in the
neutral pion decays are much above the energy corresponding to the
maximum of the pair
production rate on the IC background.
For
example, for the $10^{17}$~eV \gr s the 
cross-section of the pair production on the IC background
is $s/\mbox{ln}s\sim
10^4$ times smaller than the peak value, 
so that it is of the order of $p\gamma$
interaction cross-section. Thus, the mean free path of the secondary
\gr s is comparable to the mean free path of the primary protons and
is of order of the size of the IC radiation region.
These \gr s can give rise to electromagnetic cascades redistributing
energy into TeV \gr s. The drawback of this
mechanism of TeV emission is its low efficiency \citep{neronov03} 
due to the fact that
the TeV \gr s which are observed at infinity can be produced only in
the surface layer of the 
IC radiation region, where the optical depth for the
TeV radiation drops to the values of order one.  
On the other hand, this mechanism should result in strong multi-GeV
emission which can escape through the whole inverse
Compton radiation region without significant absorption. 
This implies that if this mechanism is indeed responsible for the TeV
emission in (some of) the TeV blazars, it should lead to a strong
signal in the multi-GeV band accessible for the Fermi (GLAST)
satellite\footnote{The official website of the Fermi collaboration is http://fermi.gsfc.nasa.gov/}. 

Apart from the interactions with soft radiation background, protons can
also interact with the matter in the jet. The optical depth of
protons with respect to this process is estimated as
\begin{equation}
\tau_{pp}=\sigma_{pp}n_p(R)R\;,
\end{equation}
where $n_p(R)$ is the matter density in the jet at the distance $R$.
Taking the $pp$ interaction cross section at energies $E_p\sim
10^{18}$~eV to be equal to
$\sigma_{pp}\simeq 10^{-25}$~cm$^2$ we obtain
\begin{equation}
\label{taupp}
\tau_{pp}\simeq 0.1
\left[\frac{n_p(R)}{10^8\mbox{ cm}}\right]\left[\frac{R}{10^{16}\mbox{ cm}}\right]\;.
\end{equation}
Particle multiplicity in $pp$ collisions at $E_p \sim 10^{18}$ eV is
about $N \sim 100-200$ (see e.g. \citep{multi} for a review). 
Thus every
collision of 
a $10^{18}$ eV proton with a 
background proton produces $\sim 100$ photons with
$E_\gamma = 5 \times 10^{15}$ eV from $\pi^0$, $\sim 100$ neutrinos
with the same energy from $\pi^{\pm}$ decays and $\sim 100$
electrons and positrons. 
All neutrinos escape from the interaction
region, while electrons, positrons and photons give rise to an
electromagnetic cascade. As we discuss in the next section, this
provides a link with the standard picture of VHE emission by
relativistic blobs in the jet. The limitation of this mechanism is
that it requires rather large densities of matter in the jet, see
Eq.~(\ref{taupp}). 

To summarize, the qualitative analysis of 
this subsection shows that production of
VHE \gr s can, in principle, be possible within the compact source
model with proton acceleration, see Fig. 1, lower panel. Still, the details of the mechanism of conversion of the proton flux power  into the power of TeV \gr\ emission are yet to be worked out.

\subsection{Particle cascade in the accretion flow: a link to the
  "relativistic blob in the jet" picture}
\label{blobs}

The $\gamma\gamma$, $p\gamma$ and $pp$ interactions taking place
during the propagation of the high-energy particles through the RIAF
environment give
rise to electromagnetic cascades, which redistribute the power
initially contained in the highest energy particles to the lower
energy bands. This is expected to result 
in the broad-band (radio-to-\gr)
electromagnetic emission from the cascade.

To large extent, this emission can be described by the conventional
synchrotron -- self Compton and/or synchrotron -- external Compton
models based on the picture of relativistically moving blobs of plasma
in the jet. Indeed,  as already mentioned, the primary flow of high-energy particles, accelerated by the compact source, is expected to be highly anisotropic, with the direction set by the magnetic field in the acceleration region.   The cascades produce a stream of relativistic
particles with velocities scattered within some angle $\zeta$ around
the direction of the primary flow. From the
kinematical point of view such a stream is nothing else than a blob of
plasma moving with the bulk Lorentz factor
\begin{equation}
\Gamma\simeq \frac{1}{\zeta}\;.
\end{equation}
The angle $\zeta$ is determined by the dynamics of the cascade. An
important role in this dynamics is played by the value and
configuration of the magnetic field. At the late stages of the cascade
development, when the density of the particles in the cascade is
large, the problem should be solved self-consistently taking into
account the back-reaction of the plasma in the cascade on the magnetic
field. This observation makes the link between the cascade and the
blob pictures not only kinematical but also dynamical: in the standard
approach the magnetic field is also determined self-consistently by
the dynamics of the blob itself.

Existence of this link enables to use in the analysis of TeV \gr\
flares within the framework of the compact source models many
results from the standard approach. In particular, the transparency of
the blob for the TeV radiation in the July 2006 large flare of \pks\
implies the constraint \citep{begelman07} $\Gamma\gtrsim 50$ and
hence, $\zeta\lesssim 1.1^\circ$.

On the other hand, there are 
several important differences between the cascade and a generic
relativistic blob model which potentially enable to distinguish the two
models observationally. First, in the cascade setup, the
relativistic blobs are formed in a compact region close to the black
hole. In this way this setup naturally incorporates the observed fast
variability and is not in a direct conflict with the low bulk Lorenz factors
observed at the parsec-scale distances.  Indeed, the bulk motion of the secondary particles in the cascade can decelerate at  parsec-scale distances either because of development of intrinsic instabilities or because of interaction with the interstellar medium. To study this possibility, a dynamical numerical calculation of the propagation of the high-energy particle cascade through the accretion flow and through the interstellar medium is needed. Second, the cascade transfers
power from the higher-energy particles to the lower-energy ones: in
this sense it is a "top-down" scenario of formation of the spectrum
of emitting electrons. It would be interesting to understand if 
the latter property 
can explain the existence of
the low-energy cut-offs in the spectra of electrons in the 
synchrotron -- inverse Compton scenarios \citep{krawchinsky07}.
Finally, as already mentioned, in the compact source model, the
emission from the blob may be superimposed onto the \gr\ component
originating directly in the acceleration region near the black hole. 

\section{Timing of the supermassive black hole 
in \pks\ with TeV $\gamma$-rays}
\label{TIMING}

Within the compact source model the spectral and variability
properties of the VHE signal can be interpreted in a qualitatively
different way (as compared to the model of VHE emission from the
parsec-scale distances). They can be used to ``probe'' the physical
environment in the central engine: the matter density, the radiative
efficiency of the accretion flow, the magnetic field
(cf. \citet{aharonian08a}). 
Moreover, in
this case the VHE observations provide a completely new tool for the
study of the black hole physics. Namely, the \gr\ timing data can be
used to derive constraints on the parameters of the black hole
itself. Below we explore such a possibility on a particular example of
the July 2006 large flare of \pks\ (redshift $z=0.116$) reported by
\citep{aharonian07}.

\subsection{The relevant time scales}

Timing properties of emission produced in the vicinity of black hole
horizon can be characterized by several fundamental time
scales. First, the minimal possible variability time scale is
determined by the requirement of causal connection of the emission
region and is given by the light crossing time of the black hole horizon,
\begin{eqnarray}
\label{lc}
t_{\rm lc}&=&2\left(R_g+\sqrt{R_g^2-a^2}\right)/c\\
&\simeq&\left\{
\begin{array}{ll}
10^2\left[M_{\rm BH}/10^7M_\odot\right]\mbox{ s,}&a=R_g\\
2\times 10^2\left[M_{\rm BH}/10^7M_\odot\right]\mbox{ s,}&a=0
\end{array}
\right.\nonumber
\end{eqnarray} 
where $R_g$
is defined in Eq.~(\ref{Rg}); the parameter $a$ is related to 
the angular
momentum $J_{\rm BH}$ of the black hole as
$a=J_{\rm BH}/M_{\rm BH}c^2$ and lies in the range $0\le a\le R_g$.

If the \gr\ emission is produced close to the black hole, variability
at the characteristic time scale of rotation around the black hole is
expected on general grounds, unless the entire system (the accretion
flow with an embedded particle acceleration region) is perfectly
axially symmetric. The intensity of the modulation of the signal can
depend on various parameters, such as the inclination angle of the
observer with respect to the BH rotation axis, distance of the
emission region from the horizon, etc. A perfect axial symmetry, which
would wash out the modulation of the signal with the period of
rotation around the black hole, can be expected in a stationary, quiet
state of the source. On the contrary, a bright flare is, most
probably, related to rapid change of the system parameters  (for  example, inspiralling of a denser clump of matter into the black hole) which
should lead to a significant disturbance of the axially symmetry.

Close to the horizon of a spinning black hole, the accreting matter
rotates around the black hole with the period \citep{bardeen}
\begin{equation}
\label{eq:P}
P(r)=2\pi\frac{r^{3/2}\pm aR_g^{1/2}}{cR_g^{1/2}}\;,
\end{equation}
where  $r$ is the radius of
the orbit. The $+$ ($-$) sign corresponds to the prograde (retrograde) orbit.  It is known that  stable circular
orbits exist only down to certain distance $r_{\rm ms}$ from the
BH. The period of rotation along the last prograde stable orbit at the
distance $r_{\rm ms}$ is
\begin{equation}
\label{eq:period}
P(r_{\rm ms})=\left\{
\begin{array}{ll}
  4\pi R_g/c\simeq 630\left[\frac{M_{\rm BH}}{10^7M_\odot}\right]
\mbox{ s,}&a=R_g\\
  12\sqrt{6}\pi R_g/c\simeq 4600
\left[\frac{M_{\rm BH}}{10^7M_\odot}\right]\mbox{ s,}&a=0
\end{array}
\right.
\end{equation}
The disturbance of the axial symmetry of the accretion flow is
expected to result in the modulation of physical conditions in the AGN
central engine with a period given by Eq.~(\ref{eq:period}).
Since the properties of the \gr\ emission from the
central engine (from the base of the jet) are determined by the
physical conditions in the central engine, the modulation of these conditions should result in the modulation of the \gr\ signal
with the same period.

\subsection{Fitting the lightcurve of \pks}

\begin{figure}
\begin{center}
\includegraphics[width=\linewidth]{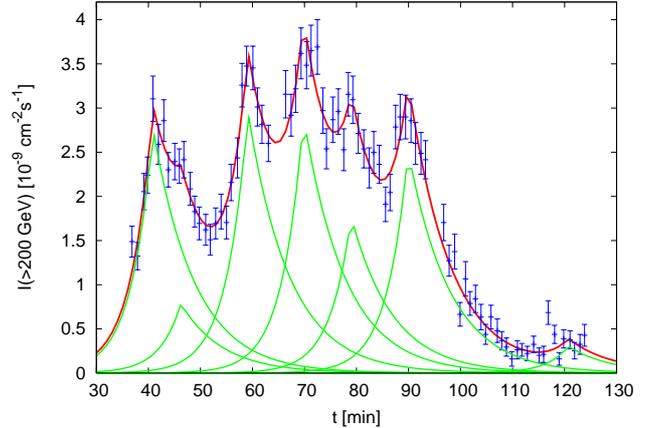}
\caption{The lightcurve of PKS 2155-304 fitted with a sequence of
  sub-flares with identical profiles,
 but with different
  normalizations (see Table \ref{tab:subflares}).}
\label{fig:PKS_lc}
\end{center}
\end{figure}

The VHE \gr\ lightcurve of the flare of \pks, reported in
\citep{aharonian07}, consists of several pronounced sub-flares (see
Fig. \ref{fig:PKS_lc}). At least three characteristic time scales can
be found in a straightforward way from the analysis of the 
lightcurve: the rise and decay times of the individual sub-flares,
$t_{\rm rise},t_{\rm decay}$, and the period of recurrence of the
sub-flares, $T$. In order to estimate the average values
of these parameters we have
fitted the VHE lightcurve assuming that the time profiles of the
individual sub-flares are characterized by the same rise and decay
times, and differ only in the arrival times and the 
amplitudes\footnote{This model is different from the model considered by
\citep{aharonian07} where the rise and decay times were allowed to vary
among individual sub-flares.}.
The approximation of identical $t_{\rm rise}$, $t_{\rm decay}$ appears
naturally in the context of the compact source models where
these parameters are determined by the intrinsic time 
scales of the central engine, such as e.g. the light crossing time of the black
hole and the cascade development 
time.
On the other hand, it is clearly a strong idealization: the
complicated dynamics of the particle acceleration and the cascade
development is expected to introduce a scatter in the
characteristics of the individual sub-flares. Still, 
we stick to this approximation as a natural
first step.

The individual sub-flares are modeled with the profile
\begin{equation}
\label{eq:profile}
I_k(t)= \begin{cases}
N_k\exp[(t-t_{{\rm max},k})/t_{\rm rise}]~,&t<t_{{\rm max},k} \\
N_k\exp[-(t-t_{{\rm max},k})/t_{\rm decay}]~,&t>t_{{\rm max},k}
\end{cases}
\end{equation}
where the time $t_{{\rm max},k}$ corresponds to the maximum intensity
and $N_k$ is the amplitude of the $k$-th sub-flare. Note that we
define $t_{\rm rise}$ ($t_{\rm decay}$) as the time in which the
signal increases (decreases) by a factor $e$. We fit the lightcurve
with the sum of several sub-flares (\ref{eq:profile}) and a constant
signal. The result of the fit of the overall lightcurve with such a
model is shown in Fig.~\ref{fig:PKS_lc}. The rise and decay times
inferred from the fit are
\begin{align}
&t_{\rm rise}=(2.5\pm 0.2)\times 10^2\mbox{ s},
\label{tr}\\
&t_{\rm decay}=(4.9 \pm 0.5)\times 10^2\mbox{ s},
\label{td}
\end{align}
The arrival times $t_{{\rm max},k}$ and
normalizations $N_k$ of the sub-flares are summarized
in the Table~\ref{tab:subflares}.

\begin{table}
\begin{tabular}{|l|l|l|l|}
\hline
k&$t_{{\rm max},k}[\mbox{min}]$ &$N_k[10^{-9}\mbox{cm}^{-2}\mbox{s}^{-1}]$ \\
\hline
1&$40.9\pm 0.3$&$2.8\pm 0.3$\\
2&$46.2\pm 0.8$&$0.78\pm 0.24$\\
3&$59.1\pm 0.2$&$3.0\pm 0.17$\\
4&$69.8\pm 0.3$&$2.9\pm 0.2$\\
5&$79.0\pm 0.4$&$1.8\pm 0.2$\\
6&$89.9\pm 0.3$&$2.5\pm 0.2$\\
9&$120.7\pm 1.0$&$0.30\pm 0.11$\\
\hline
\end{tabular}
\caption{List of parameters of the sub-flares.}
\label{tab:subflares}
\end{table}

Initially, we fitted the lightcurves with a set of five
sub-flares which correspond to the five pronounced peaks clearly
visible in the data. The quality of this fit was rather low 
($\chi^2=106$ for 70
degrees of freedom). We have found that the addition of one more
sub-flare (sub-flare number 2 in Table~\ref{tab:subflares})
significantly improves the quality of the fit ($\chi^2=86.3/68$
d.o.f.; the F-test 
gives a chance probability of the fit improvement at
the level of 0.1\%).  Finally, we have found that the fit can be
further improved by addition of one more sub-flare near the end of the
overall lightcurve, the last sub-flare in the Table
\ref{tab:subflares}. The quality of the fit with seven sub-flares is
$\chi^2=75.8/66$ d.o.f.  According to the F-test the probability that
the latter fit improvement is achieved by chance is $1.4~\%$.

\begin{figure}
\begin{center}
\includegraphics[width=\linewidth]{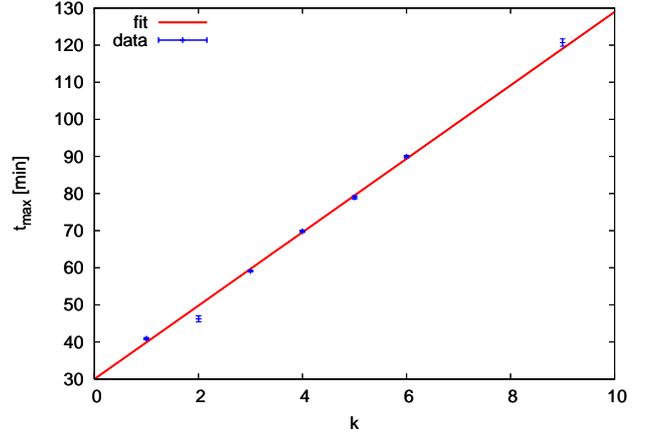}
\caption{The arrival times of the sub-flares of the PKS2155-304
  lightcurve as a function of the sub-flare number. 
The data are fitted by the straight line $t_{{\rm max},k}=t_0+k\cdot T$. }
\label{fig:period}
\end{center}
\end{figure}

The arrival times of the six bright sub-flares follow an
approximate linear law 
\begin{equation}
\label{eq:time}
t_{{\rm max},k}\simeq t_0+k\cdot T\;,
\end{equation}
see Fig.~\ref{fig:period}. This is reminiscent of quasi-periodic
oscillations observed in the X-ray band in the X-ray binaries
\citep{2000ARA&A..38..717V} and in the infrared band in the Galactic Center
\citep{2003Natur.425..934G}.  The last weak sub-flare also falls on
the linear dependence (\ref{eq:time}) if assigned the number
$k=9$. However, we do not use this sub-flare in the following analysis
because its presence is not strongly required by the lightcurve fit.

Fitting the set of arrival times $t_{{\rm max},k}$ of the six bright 
sub-flares (numbered by 1--6 in the Table~\ref{tab:subflares}) with
the linear law (\ref{eq:time}) one finds the recurrence
period of the sub-flares, 
\be
T=(5.93\pm 0.15)\times 10^2\mbox{ s}\;.
\label{eq:rd} 
\ee
This recurrence
period coincides with the characteristic variability scale
$\sim 600$~s mentioned by \citet{aharonian07}.

We have tested
if the apparent quasi-periodicity of the signal can appear in the data
by chance.  To do this we have fixed the arrival times of the first
and last of the bright sub-flares (sub-flares 1 and 6 in the Table
\ref{tab:subflares}). Then we have allowed the arrival times of the 4
intermediate sub-flares to be distributed randomly between 
$t_{{\rm max},1}$ and $t_{{\rm max},6}$. The arrival times $t_{{\rm
    max},k}$, $k=1,\ldots,6$ are assigned the errors from the
Table~\ref{tab:subflares}. We have calculated the probability that
fitting the arrival times of the six first sub-flares by the function
(\ref{eq:time}) would result in a $\chi^2\le \chi_0^2$, where
$\chi_0^2=41$ is the $\chi^2$ of the fit of 
the real data\footnote{Note that $\chi_0^2$ is large. This means
  that the oversimplified model of strictly identical sub-flares with
  strictly periodic 
  arrival times is actually excluded by the data. This is not
  surprising:
as already mentioned,
on physical grounds one expects deviations
  from this idealization.}.
This
probability turns out to be $1.3\cdot 10^{-3}$.  We also tried another
error assignment. Namely, all the arrival times were assigned the
error $\delta t=1.5$~min, which corresponds to the intrinsic scatter
of the arrival times of the real sub-flares around the linear law
(\ref{eq:time}). In this case the best linear fit corresponds to
$\chi^2_1=5.43$ for 4 degrees of freedom. The chance probability to
obtain $\chi^2\le 5.43$ in the simulated data sets with random arrival
times of the intermediate sub-flares turns out to be $6\cdot 10^{-3}$
in this case. 
We have checked that our conclusions do not depend on the particular
choice (\ref{eq:profile}) of the sub-flare profile by
considering other possible choices of $I_k(t)$.

\subsection{Constraints on the black hole parameters}
\label{sec:3.3}

\begin{figure}
\begin{center}
\includegraphics[width=\linewidth]{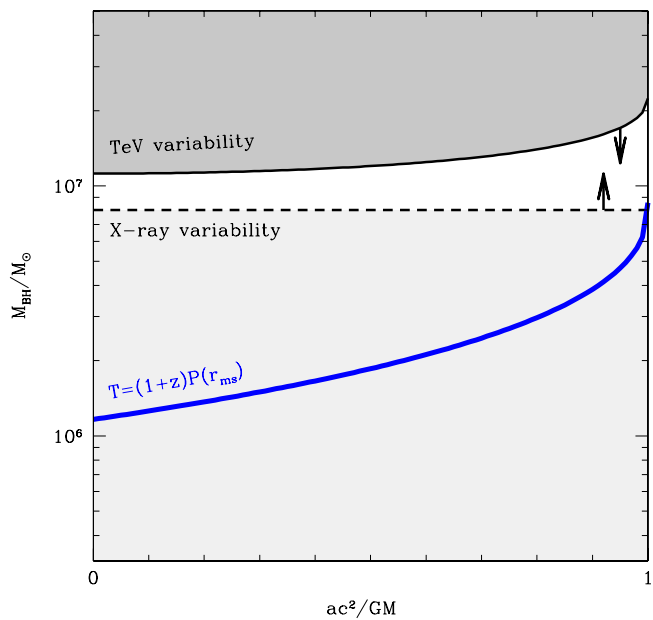}
\caption{The parameters of the supermassive black hole in \pks\
  inferred from the VHE \gr\ lightcurve. The shaded region shows the
  range of $M_{\rm BH}$, $a$ excluded by the requirement
  (\ref{riselc}). 
Thick blue curve shows the relation between
  $M_{\rm BH}$ and $a$ obtained by identifying the period of the sub-flare
  recurrence with the rotation period over the last stable orbit.
  The horizontal line shows the lower bound on the black hole mass
  implied by the X-ray variability analysis of
  \citep{zhang05}. For
  $a>0$ this bound should be taken with caution as its 
dependence on $a$ has not been 
explored.}
\label{fig:period_aM}
\end{center}
\end{figure}

The rise and decay times (\ref{tr}), (\ref{td}) as well as the
recurrence time (\ref{eq:rd}) are determined by the physics of the
\gr\ emission. Within the compact source scenario 
described in Sec.~\ref{IR}, the minimal
possible rise time is given by the light crossing time of the black
hole, Eq. (\ref{lc}). Requiring
\begin{equation}
\label{riselc}
t_{\rm rise}\ge (1+z)t_{\rm lc}
\end{equation}
 one finds 
a range of parameters  $M_{\rm BH}, a$ excluded by
the observations. This range corresponds to the dark shaded region  in Fig. \ref{fig:period_aM}.

This should be compared with the constraints obtained
by other methods. In the literature there are two estimates of the
mass of the central black hole in \pks.

The first one \citep{aharonian07} is obtained from the relation between
the masses of central black holes and luminosities of host galaxy bulges
\citep{bettoni03}; it gives $M_{\rm BH}\sim 10^9 M_\odot$.  As it was
already noted in \citep{aharonian07}, this estimate is in conflict with
the constraint $M_{\rm BH}\le 2\times 10^7M_\odot$ derived from the
\gr\ variability properties. We point
out, however, that there are several uncertainties in the estimate 
based on the $M_{\rm BH}-L_{\rm
  bulge}$ relation. It is obtained by
extrapolating the empirical relation observed in a local sample of normal
galaxies to the case of TeV blazars. The reliability of this
extrapolation was never investigated. Besides, the relation itself has 
a large intrinsic scatter (order-of-magnitude deviations are
present). 
Finally, there is no consensus in the
literature about the luminosity of the host galaxy in 
\pks\ (see \citep{aharonian07} and references therein).

The second estimate \citep{zhang05} is based on the study of X-ray
variability properties of \pks: it uses the method of
\citep{nikolajuk04} to relate the mass of black hole to the excess
variability $\sigma^2_{\rm nxs}$ at a certain frequency scale. This
method yields the bound $M_{\rm BH}\ge 8.1\times 10^6M_\odot$, shown
by a dashed horizontal line in Fig. \ref{fig:period_aM}. The latter
estimate is compatible with the constraint derived from the TeV
variability of the source.  One should note, however, that the
estimate of the black hole mass based on the X-ray variability
properties suffers from the same uncertainty as the estimate based on
the $M_{\rm BH}-L_{\rm bulge}$ relation: it is originally derived for
a sample of nearby non-blazar AGNs. Its applicability to the sample of
the TeV blazars was never tested.

 The tight constraint on the black hole mass, derived from the TeV variability time scale,
provides a possibility of strong observational test of the compact source model. Indeed, a precise determination of the black hole mass by another method would be able to falsify or confirm the compact source model. 

The indication of the quasi-periodicity of the
sub-flare arrival times suggests to associate the period $T$ of the
sub-flare recurrence with the (minimal possible) period
of rotation around the black hole,
\begin{equation}
\label{p=t}
T=(1+z)P(r_{\rm ms}).
\end{equation}
This gives a relation between $a$ and $M_{\rm BH}$ shown by the thick
blue curve in Fig.  \ref{fig:period_aM}. If combined with the
constraint on the black hole mass derived from the X-ray variability
analysis \citep{zhang05}, this relation implies that the black hole is
rotating almost at the maximal rate, $a\approx R_g$. 
However, we remind 
that the X-ray constraint on the black hole mass should be
taken with caution. In particular, to the best of our knowledge, the
dependence of this phenomenological constraint on the black hole spin
$a$ has not been explored.

\section{Conclusions}
\label{CONCL}

In this paper we have proposed that the recently observed fast
variability of the VHE emission from blazars can be naturally
accommodated within the framework of "compact source" model. In this
model particles responsible for the observed VHE emission are
accelerated close to the central supermassive black hole, rather than
at large distances downstream the AGN jet.  We have analyzed the
problem of escape of the VHE \gr s from the vicinity of the central
engine and demonstrated that the region around the central engine is
transparent for TeV \gr s if the accretion flow in the TeV blazars is
radiatively inefficient. If the luminosity of the accretion flow is as
low as $L_{\rm acc}\lesssim 10^{40}\mbox{erg/s}$, the TeV \gr\
emission can come directly from the immediate neighborhood of the
central black hole. Alternatively, for brighter accretion flows, the
TeV \gr s may be produced at some distance from the black hole in a
proton-initiated electromagnetic cascade developing in the accretion
flow.

The possibility that the properties of the VHE \gr\ emission are
directly linked to the properties of the central engine of the AGN, if
confirmed by future observations, provides a new tool to study the
physical conditions in the direct vicinity of the supermassive black
hole. In particular, the VHE signal can be used to constrain the
parameters of the accretion flow and of the black hole itself, such as
its mass and spin.

We demonstrated this possibility on the example of the bright TeV
flare of the blazar \pks. Within the proposed scenario, the
characteristic time scales, found in the timing analysis of this flare,
are directly related to the parameters of the supermassive black
hole. The minimal variability time scale of the signal is identified
with the black hole light-crossing time. This sets the bound on the black
hole mass and its rotation moment shown in Fig.~\ref{fig:period_aM}.
We also observed that the signal exhibits quasi-periodic
oscillations. 
Identifying the
recurrence time of these oscillations 
with the period of rotation
around the black hole we obtained a relation between the black hole
mass and its rotation moment.  

A detailed modeling based of the framework proposed in this paper
should involve calculation of particle
acceleration and propagation in the vicinity of the black hole through
the environment created by the accretion flow. This can be done 
assuming particular (numerical) models of RIAF 
and particle
acceleration. We leave this for future work.

\subsection*{Acknowledgments}
We thank F.~Bezrukov, A.~Boyarsky, G.~Dvali, D.~Horns, K.~Postnov, V.~Rubakov,
O.~Ruchayskiy, G.~Sigl, P.~Tinyakov and I.~Tkachev
for useful discussions and comments. The work of S.S. was partially
supported by the EU 6th Framework Marie Curie Research and Training
network "UniverseNet" (MRTN-CT-2006-035863). D.S. thanks Theoretical
Department of CERN for hospitality during initial stages of this work.

\appendix

\section{Synchrotron  self-absorption in the accretion
  flow} 
\label{App}

The function $I(x)$ entering into the expression (\ref{alphaSA}) for
the synchrotron self-absorption coefficient has the form
\citep{Pacholczyk},
\be
I(x)=\frac{1}{x}\int_0^\infty z^2\e^{-z}F(x/z^2)dz\;,
\ee
where 
$
F(x)=x\int_x^\infty K_{5/3}(z)dz
$.
It is straightforward to obtain the asymptotics of $I(x)$,
\be
\label{Iasymp}
I(x)\sim \sqrt{2/3}\pi\exp\big(-3(x/4)^{1/3}\big)~,~~~~x\gg 1\;.
\ee

The synchrotron photons are self-absorbed below a certain energy
$\epsilon_{\rm SA}$. The latter is estimated from the condition that
the optical depth for the synchrotron emission with energy $\epsilon_{\rm SA}$ is equal
to unity,
\be
\alpha_{\rm SA}(\epsilon_{\rm SA})R_{\rm synch}=1\;,
\ee 
where $R_{\rm synch}$ is the size of the synchrotron emission
region. Taking $R_{\rm synch}\approx R_g$ one obtains the following
equation for the variable\footnote{See Eq. (\ref{x}) for the
  definition of this variable.} 
$x_{\rm SA}$ corresponding to the
self-absorption energy $\epsilon_{\rm SA}$,
\be
\label{xeq}
\begin{split}
&1.89 x_{\rm SA}^{1/3}+\ln{x_{\rm SA}}
=17.37\\
&+\ln\left\{\!
\bigg[\frac{n}{10^{10}\,\mbox{cm}^{-3}}\bigg]\!
\left[\frac{M_{\rm BH}}{10^7 M_\odot}\right]\!
\left[\frac{B}{10^4\mbox{ G}}\right]^{-1}\!
\left[\frac{T_e}{1\mbox{ MeV}}\right]^{-5}\!
\right\}\;.
\end{split}
\ee
When the logarithm in the second line vanishes, the solution to this
equation is 
\be
\label{xval}
x_{\rm SA}=247\;.
\ee
This corresponds to the value (\ref{SA}) of the self-absorption
energy. Note that the r.h.s. of the equation (\ref{xeq}) 
logarithmically depends on
the parameters of the accretion flow implying
that the variable $x_{\rm SA}$ is only mildly sensitive to these
parameters. Namely, the value (\ref{xval}) 
is multiplied by a factor ranging
from $0.6$ to $1.5$ when the combination of the parameters entering
the logarithm in the second line of Eq.~(\ref{xeq}) varies from $0.1$
to $10$.

 \label{lastpage}

\end{document}